\documentclass[aps,pre,twocolumn,amsmath,amssymb,showpacs,showkeys]{revtex4}
\usepackage{graphicx}% Include figure files
\usepackage{epsfig}
\usepackage{dcolumn}% Align table columns on decimal point
\usepackage{bm}% bold math

\begin{document}
\title{An outlook on correlations in stock prices}

\author{Anirban Chakraborti}%
  \email{achakraborti@yahoo.com}
  %\homepage{http://www.cmth.bnl.gov/~anirban/}
  \affiliation{Department of Physics, Banaras Hindu University,
  Varanasi-221005, India}

\date{\today}

\begin{abstract}
We present an outlook of the studies on correlations in the price time-series 
of stocks, discussing the construction and applications of "asset tree".
The topic discussed here should illustrate how the complex 
economic system (financial market) enrichens the list of existing dynamical 
systems that physicists have been studying for long.
\end{abstract}


\maketitle

%\tableofcontents

\graphicspath{{./fig/}}  %% PATH FOR FIGURES

\begin{quote}
{\it ``If stock market experts were so expert, they would be buying stock, not selling advice.''} \\
-- Norman Augustine, US aircraft businessman (1935 - )
\end{quote}
\section{Introduction}

The word ``correlation'' is defined as ``a relation existing between phenomena 
or things or between mathematical or statistical variables which tend to vary, 
be associated, or occur together in a way not expected on the basis of chance 
alone'' (see http://www.m-w.com/dictionary/correlations). As soon as we talk 
about ``chance'', the words ``probability'',``random'', etc come to our mind. 
So, when we talk about correlations in stock prices, what we are really 
interested in are the nature of the time series of stock prices, the relation 
of stock prices with other variables like stock transaction volumes, the 
statistical distributions and laws which govern the price time series, in 
particular whether the time series is random or not. The first formal efforts
in this direction were those of Louis Bachelier, more than a century ago 
\cite{Bachelier}. Eversince, financial time 
series analysis is of prevalent interest to theoreticians  for 
making inferences and predictions though it is primarily an empirical 
discipline. The uncertainty in  the financial time series and its theory makes 
it specially interesting to statistical physicists, besides 
financial economists \cite{tsay,santha}.
One of the most debatable issues in financial economics is whether the market 
is ``efficient'' or not. The ``efficient'' asset market is one in which the 
information 
contained in past prices is instantly, fully and continually reflected in the 
asset's current price. As a consequence, the more efficient the market is, the 
more random is the
sequence of price changes generated by the market. Hence, the most efficient
market is one in which the price changes are completely random and 
unpredictable. This leads to another relevant or pertinent question of 
financial 
econometrics: whether asset prices are predictable. Two simplest models of 
probability theory and financial econometrics that deal with predicting 
future price changes, the random walk theory and Martingale theory, assume that
the future price changes are functions of only the past price changes. 
Now, in Economics the ``logarithmic returns'' is calculated using the formula
\begin{equation}
r(t)=\ln P(t)-\ln P(t-1),
\label{ret}
\end{equation}
where $P(t)$ is the price (index) at time step $t$.
A main
characteristic of the random walk and Martingale models is that the 
returns are {\it uncorrelated}.

\begin{figure}[]
\centering
\includegraphics[width=2.5in,angle=-90]{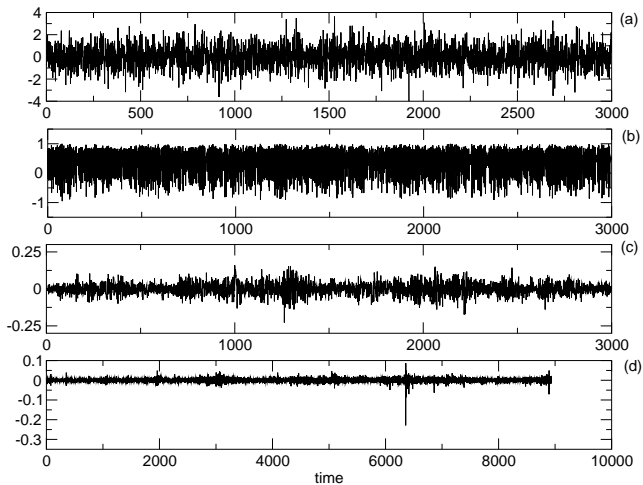}
\caption{
Comparison of several time series which are of interest to physicists and 
economists: (a) Random time series (3000 time steps) using random numbers from
a Normal distribution with zero mean and unit standard deviation. 
(b) Multivariate 
spatio-temporal time series (3000 time steps) drawn from the class of 
diffusively 
coupled map lattices in one-dimension with sites $i=1,2...n'$ of the form: 
$y_{t+1}^i = (1-\epsilon) f(y_t^i) + \frac{\epsilon}{2}( f(y_t^{i+1})
+ f(y_t^{i-1}))$,
where $f(y) = 1 - a y^2$ is the logistic map whose dynamics is
controlled by the
parameter $a$, and the parameter $\epsilon$ is a measure
of coupling between nearest-neighbor lattice sites. We use parameters 
$a=1.97$, $\epsilon=0.4$ for the dynamics to be in the regime of spatio-temporal chaos. We choose $n=500$ and iterate, 
starting from random initial conditions, for $p=5 \times 10^7$
time steps, after discarding $10^5$ transient iterates. Also, we choose
periodic boundary conditions, $x(n+1)=x(1)$.
(c) Multiplicative stochastic process GARCH(1,1) for a random variable $x_{t}$
with zero mean and variance $\sigma_t^2$, characterized by a Gaussian 
conditional probability distribution function $f_t(x)$:
$\sigma_t^2=\alpha_0+\alpha_1 x_{t-1}^{2}+\beta_{1}\sigma_{t-1}^{2}$,
using parameters $\alpha_0 = 0.00023$, $\alpha_1 = 0.09$ and
$\beta_1 = 0.01$ (3000 time steps). (d) Empirical Return time series of the 
S\&P500 stock index (8938 time steps).
}
\label{fig1}
\end{figure}

In the past, several hypotheses have been proposed to model financial time 
series and studies have been conducted to explain their most characteristic 
features. 
The study of long-time correlations in the financial time series is a very 
interesting 
and widely studied problem, especially since they give a deep insight about the
underlying processes that generate the time series \cite{stanley}. The complex 
nature of financial time series (see  Fig.~\ref{fig1}) has especially forced the 
physicists to add this system to their existing list of dynamical systems that 
they study.
Here, we will not try to review
all the studies, but instead give a brief outlook of the studies done by the
author and his collaborators, and the motivated readers are kindly asked to 
refer the original papers for further details.

\section{Analysing Correlations in Stock Price time series}

%%%%%%%%%%%%%%%%%%%%%%%%%%%%%%%%%%%%%%
%%%%%%%%%%%%%%%%%%%%%%%%%%%%%%%%%%%%%%

\subsection{Financial Correlation matrix and constructing Asset Trees}

% Data used in the study
In our studies, we used two different sets of financial data for 
different purposes. The first set from the Standard \& Poor's 500 index 
(S\&P500) of the New York Stock Exchange (NYSE) from July 2, 1962 to 
December 31, 1997 containing 8939 daily closing values, which we have already 
plotted in Fig.~\ref{fig1}(d).
In the second set, we study the split-adjusted daily closure prices 
for a total of $N=477$ stocks traded at the New York Stock Exchange (NYSE) 
over the period of 20 years, from 02-Jan-1980 to 31-Dec-1999. This 
amounts a total of 5056 price quotes per stock, indexed by time variable 
$\tau = 1, 2, \ldots, 5056$. For analysis and smoothing purposes, the 
data is divided time-wise into $M$ \emph{windows} $t=1,\, 2,...,\, M$ 
of width $T$, where $T$ corresponds to the number of daily returns included in 
the window. Several consecutive windows overlap with each other, the 
extent of which is dictated by the window step length parameter 
$\delta T$, which describes the displacement of the window and is also measured  
in trading days. The choice of window width is a trade-off between too 
noisy and too smoothed data for small and large window widths, 
respectively. The results presented in this paper were calculated from 
monthly stepped four-year windows, i.e. $\delta T = 250/12 \approx 21$ days and 
$T=1000$ days. We have explored a large scale of different values for both 
parameters, and the cited values were found optimal \cite{jpo}. With 
these choices, the overall number of windows is $M=195$.

% Correlations
In order to investigate correlations between stocks we first denote 
the closure price of stock $i$ at time $\tau$ by $P_{i}(\tau)$ 
(Note that $\tau$ refers to a date, not a time window). We focus 
our attention to the logarithmic return of stock $i$, given by 
$r_{i}(\tau)=\ln P_{i}(\tau)-\ln P_{i}(\tau-1)$ which for a sequence 
of consecutive trading days, i.e. those encompassing the given window 
$t$, form the return vector $\boldsymbol r_{i}^t$. In order to 
characterize the synchronous time evolution of assets, we use the equal 
time correlation coefficients between assets $i$ and $j$ defined as

\begin{equation}
\rho _{ij}^t=\frac{\langle \boldsymbol r_{i}^t \boldsymbol r_{j}^t \rangle -\langle \boldsymbol r_{i}^t \rangle \langle \boldsymbol r_{j}^t \rangle }{\sqrt{[\langle {\boldsymbol r_{i}^t}^{2} \rangle -\langle \boldsymbol r_{i}^t\rangle ^{2}][\langle {\boldsymbol r_{j}^t}^{2} \rangle -\langle \boldsymbol r_{j}^t \rangle ^{2}]}},
\end{equation}

\noindent where $\left\langle ...\right\rangle $ indicates a time 
average over the consecutive trading days included in the return 
vectors. These correlation 
coefficients fulfill the condition $-1\leq \rho _{ij}\leq 1$. 
If $\rho _{ij}=1$, the stock price changes are completely correlated;
if $\rho _{ij}=0$, the stock price changes are uncorrelated and if
$\rho _{ij}=-1$, then the stock price changes are completely 
anti-correlated \cite{jp}.  
These correlation coefficients form an $N\times N$ correlation matrix 
$\mathbf{C}^t$, which 
serves as the basis for trees discussed in this paper.

% tree construction

We construct an asset tree according to the methodology by Mantegna 
\cite{Man1}. 
For the purpose of constructing asset trees, 
we define a distance between a pair of stocks.
This distance is associated with the edge connecting the 
stocks and it is expected to reflect the level at which the stocks are correlated.
We use a simple non-linear transformation $d^t_{ij}=\sqrt{2(1-\rho _{ij}^t)}$ to 
obtain distances with the property $2\geq d_{ij}\geq 0$, forming an 
$N\times N$ symmetric distance matrix $\mathbf{D}^t$. 
So, if $d_{ij}=0$, the stock price changes are completely correlated;
if $d_{ij}=2$, the stock price changes are completely anti-uncorrelated.
The trees for different time windows are not 
independent of each other, but form a series through time. 
Consequently, this multitude of trees is interpreted as a 
sequence of evolutionary steps of a single \emph{dynamic asset tree}.
We also require an additional hypothesis about the topology of the metric 
space, the ultrametricity hypothesis. 
In practice, it leads to determining the minimum spanning tree (MST) of the distances, denoted $\mathbf{T}^t$. 
The spanning tree is a simply connected acyclic (no cycles) graph that connects all $N$ nodes (stocks) 
with $N-1$ edges such that the sum of all edge weights, $\sum _{d_{ij}^t \in \mathbf{T}^t}d_{ij}^t$, is minimum. 
We refer to the minimum spanning tree at time $t$ by the notation $\mathbf{T}^t=(V,E^t)$, where $V$ 
is a set of vertices and $E^t$ is a corresponding set of unordered pairs of vertices, or edges.
Since the spanning tree criterion requires all $N$ nodes to be always present, the 
set of vertices $V$ is time independent, which is why the time superscript has been dropped from notation. 
The set of edges $E^t$, however, does depend on time, as it is expected that edge lengths in the matrix 
$\mathbf{D}^t$ evolve over time, and thus different edges get selected in the tree at different times.

% normalized lengths 
\subsection{Market characterization}

\begin{figure}
\epsfig{file=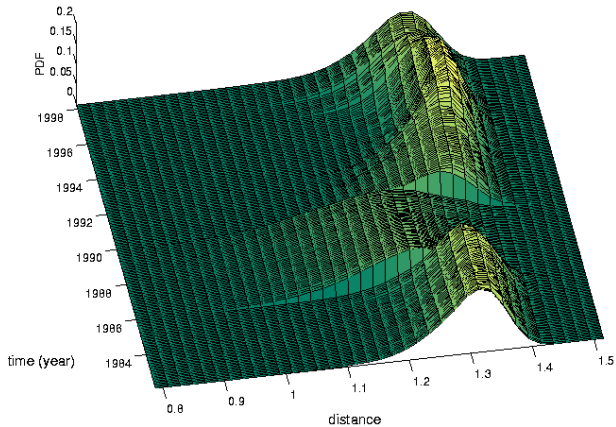,width=3.2in }
\caption{Distribution of all $N(N-1)/2$ distance elements $d_{ij}$ contained in the distance matrix $\mathbf{D}^t$ as a function of time.}
\label{all_distances}
\end{figure}

\begin{figure}
\epsfig{file=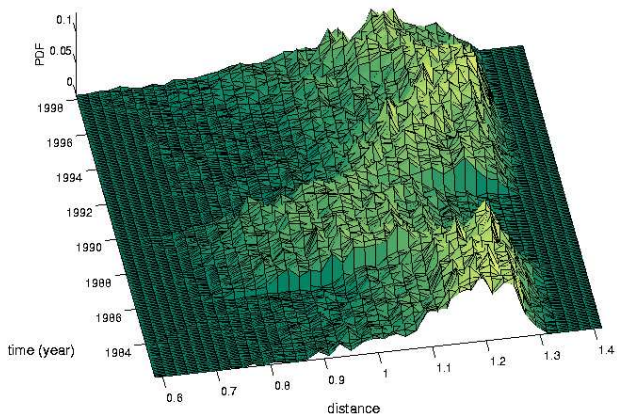,width=3.2in }
\caption{Distribution of the $(N-1)$ distance elements $d_{ij}$ contained in the asset (minimum spanning) tree $\mathbf{T}^t$ as a function of time.}
\label{mst_distances}
\end{figure}

We plot the distribution of (i) distance elements $d^t_{ij}$ contained in 
the distance matrix $\mathbf{D}^t$ (Fig.~\ref{all_distances}), (ii) distance elements $d_{ij}$ 
contained in the asset (minimum spanning) tree $\mathbf{T}^t$ (Fig.~\ref{mst_distances}). 
In both plots, but most prominently in Fig.~\ref{all_distances}, 
there appears to be a discontinuity in the distribution between roughly 1986 and 1990. The part that 
has been cut out, pushed to the left and made flatter, is a manifestation of 
Black Monday (October 19, 1987), and its length along the time axis is related to the choice of window width $T$ \cite{bali,jp}. 
Also, note that in the distribution of tree edges in Fig.~\ref{mst_distances} most edges included in the tree seem to come 
from the area to the right of the value 1.1 in Fig.~\ref{all_distances}, and the largest distance element is $d_{max}=1.3549$.

\subsubsection{Tree occupation and central vertex}

% mean occupation layer
We focus on characterizing the spread of nodes on the tree, by introducing the 
quantity of \emph{mean occupation layer}  

\begin{equation}
l(t,v_c)=\frac{1}{N}\sum _{i=1}^{N}\mathop {\mathrm{lev}}(v_{i}^{t}),\end{equation}

\noindent where $\mathop {\mathrm{lev}}(v_{i})$ denotes the level 
of vertex $v_{i}$. The levels, not to be confused with the 
distances $d_{ij}$ between nodes, are measured in natural numbers 
in relation to the \emph{central vertex} $v_c$, whose level is 
taken to be zero. Here the mean occupation layer indicates the 
layer on which the mass of the tree, on average, is conceived 
to be located.  The central 
vertex is considered to be the parent of all other nodes in the tree, and is 
also known as the root of the tree. It is used as the {\it reference }
point in the tree, against which the locations of all other nodes 
are relative. Thus all other nodes in the tree are children of 
the central vertex. Although there is an {\it arbitrariness} in the 
choice of the central vertex, we propose that it is central, 
in the sense that any change in its price strongly 
affects the course of events in the market on the whole. We have
proposed three alternative definitions for the central 
vertex in our studies, all yielding similar and, in most cases, 
identical outcomes. The idea here is to find 
the node that is most strongly connected to its nearest neighbors. 
For example, according to one definition, the central node is the one with 
the highest \emph{vertex degree}, i.e. the number of edges 
which are incident with (neighbor of) the vertex. Also, one may have either (i) static (fixed at all times) or (ii) 
dynamic (updated at each time step) central vertex, but again the results do not seem to vary significantly. We can then study the variation of the topological properties and nature of the trees, with time. This type of visualization tool can sometimes provide deeper insight of the dynamical system.

\subsubsection{Economic taxonomy}
\begin{figure}
\epsfig{file=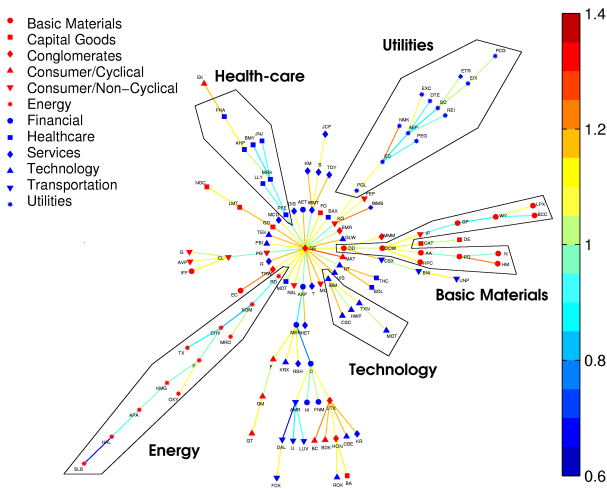,width=3.2in }
\caption{
Snapshot of a dynamic asset tree connecting the examined 116 stocks 
of the S\&P 500 index. The tree was produced using four-year 
window width and it is centered on January 1, 1998. Business 
sectors are indicated according to Forbes, \emph{http://www.forbes.com}. 
In this tree, General Electric (GE) was used as the central 
vertex and eight layers can be identified.
}
\label{samplegraph}
\end{figure}

Mantegna's idea of linking stocks in an 
ultrametric space was motivated \emph{a posteriori} by the 
property of such a space to provide a meaningful economic 
taxonomy. 
In \cite{Man1}, Mantegna examined the meaningfulness of 
the taxonomy by comparing the grouping of stocks in the 
tree with a third party reference grouping of stocks by 
their industry etc. classifications. In this case, the 
reference was provided by Forbes\cite{forbes}, which 
uses its own classification system, assigning each stock with 
a sector (higher level) and industry (lower level) category. 
In order to visualize the grouping of stocks, we constructed a 
sample asset tree for a smaller dataset (shown in Fig.~\ref{samplegraph})
\cite{short}, 
which consists of 116 S\&P 500 stocks, extending from the beginning 
of 1982 to the end of 2000, resulting in a total of 4787 price 
quotes per stock \cite{supp}. The window width was set at $T=1000$, and the 
shown sample tree is located time-wise at $t=t^*$, corresponding 
to 1.1.1998. The stocks in this dataset fall into 12 \emph{sectors}, 
which are Basic Materials, Capital Goods, Conglomerates, 
Consumer/Cyclical, Consumer/Non-Cyclical, Energy, Financial, 
Healthcare, Services, Technology, Transportation and Utilities. 
The sectors are indicated in the tree (see Fig.~\ref{samplegraph}) with different markers, 
while the industry classifications are omitted for reasons of clarity.
We use the term sector 
exclusively to refer to the given third party classification 
system of stocks. The term \emph{branch} refers to a subset of 
the tree, to all the nodes that share the specified common parent. 
In addition to the parent, we need to have a reference point 
to indicate the generational direction (i.e. who is who's parent) 
in order for a branch to be well defined. Without this reference 
there is absolutely no way to determine where one branch ends 
and the other begins. In our case, the reference is the central 
node. There are some branches in the tree, in which most of the 
stocks belong to just one sector, indicating that the branch is 
fairly homogeneous with respect to business sectors. This finding 
is in accordance with those of Mantegna \cite{Man1}, although 
there are branches that are fairly heterogeneous, such as the 
one extending directly downwards from the central vertex (see 
Fig.~\ref{samplegraph}). 

\subsection{Portfolio analysis}

Next, we apply the above discussed concepts and measures to 
the portfolio optimization problem, a basic problem of financial analysis. 
This is done in the hope that the asset tree could 
serve as another type of quantitative approach to and/or visualization 
aid of the highly inter-connected market, thus acting as a tool 
supporting the decision making process.
We consider a \emph{general Markowitz portfolio} $\mathbf{P}(t)$ with 
the asset weights $w_{1},\, w_{2},\ldots ,\, w_{N}$. In the classic
Markowitz portfolio optimization scheme, financial assets are 
characterized by their average risk and return, where the risk 
associated with an asset is measured by the standard deviation of returns. 
The Markowitz optimization is usually carried out by using historical data.
The aim is to optimize the asset weights so that the overall portfolio 
risk is minimized for a given portfolio return $r_{\mathbf{P}}$. 
In the dynamic asset tree framework, however, the task is to 
determine how the assets are located with respect to the 
central vertex. 

Let $r_{m}$ and $r_{M}$ denote the returns of the minimum and 
maximum return portfolios, respectively. The expected portfolio 
return varies between these two extremes, and can be expressed 
as $r_{\mathbf{P},\theta}=(1-\theta) r_{m} + \theta r_{M}$, 
where $\theta$ is a fraction between 0 and 1. Hence, 
when $\theta = 0$, we have the minimum risk portfolio, 
and when $\theta = 1$, we have the maximum return (maximum risk) 
portfolio. The higher the value of $\theta$, the higher the 
expected portfolio return $r_{\mathbf{P},\theta}$ and, 
consequently, the higher the risk the investor is willing to 
absorb. We define a single measure, the 
\emph{weighted portfolio layer} as

\begin{equation}
l_{\mathbf{P}}(t,\theta)=\sum _{i\in \mathbf{P}(t,\theta)}w_{i}\mathop {\mathrm{lev}}(v_{i}^{t}),\end{equation}

\noindent where $\sum_{i=1}^{N}w_i=1$ and further, as a starting point, 
the constraint $w_{i}\geq 0$ for all $i$, which is equivalent to 
assuming that there is no short-selling. The purpose of this constraint 
is to prevent negative values for $l_{\mathbf{P}}(t)$, which would 
not have a meaningful interpretation in our framework of trees with 
central vertex. This restriction will shortly be discuss further.

\begin{figure}
\epsfig{file=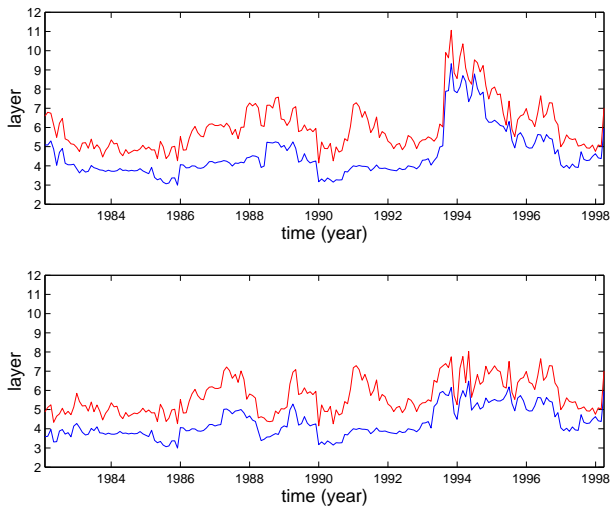,width=3.2in }
\caption{
Plot of the weighted minimum risk portfolio layer 
$l_{\mathbf{P}}(t,\theta = 0)$ with no short-selling and 
mean occupation layer $l(t,v_{c})$ against time. 
Top: static central vertex, bottom: dynamic central vertex 
according to the vertex degree criterion.
}
\label{portnss}
\end{figure}

Fig.~\ref{portnss} shows the behavior of the mean occupation 
layer $l(t)$ and the weighted minimum risk portfolio layer 
$l_{\mathbf{P}}(t,\theta = 0)$. 
We find that the portfolio layer is higher than the mean layer at 
all times. The difference between the layers depends on the window 
width, here set at $T=1000$, and the type of central vertex 
used. The upper plot in Fig.~\ref{portnss} is produced using 
the static central vertex (GE), and the difference in layers is found 
to be 1.47. The lower one is produced by using a dynamic central 
vertex, selected with the vertex degree criterion, in which case 
the difference of 1.39 is found. 
Here, we had assumed the no short-selling condition. However, it turns 
out that, in practice, the weighted portfolio layer never assumes 
negative values and the short-selling condition, in fact, is not 
necessary. 
Only minor differences are 
observed in the results between banning and allowing 
short-selling. Further, the difference in layers is also slightly larger 
for static than dynamic central vertex, although not by a significant amount.

As the stocks of the minimum risk portfolio are found on the 
outskirts of the tree, we expect larger trees (higher $L$) to 
have greater \emph{diversification potential}, i.e., the scope 
of the stock market to eliminate specific risk of the minimum 
risk portfolio. In order to look at this, we calculated the 
mean-variance frontiers for the ensemble of 477 stocks using 
$T=1000$ as the window width. If we study 
the level of portfolio risk as a function of time, we find 
a similarity between the risk curve and the curves of the mean 
correlation coefficient $\bar{\rho }$ and normalized tree 
length $L$ \cite{jp}. Earlier, when the smaller dataset of 116 stocks - 
consisting primarily important industry giants - was used, 
we found Pearson's linear correlation between the 
risk and the mean correlation coefficient $\bar{\rho}(t)$ to be $0.82$, 
while that between the risk and the normalized tree 
length $L(t)$ was $-0.90$. Therefore, for that dataset, 
the normalized tree length was able to explain the 
diversification potential of the market better than the mean 
correlation coefficient. For the current set of 477 stocks, which  
includes also less influential companies, the Pearson's linear 
and Spearman's rank-order correlation coefficients between 
the risk and the mean correlation coefficient are 0.86 and 
0.77, and those between the risk and the normalized tree 
length are -0.78 and -0.65, respectively. 

\begin{figure}
\epsfig{file=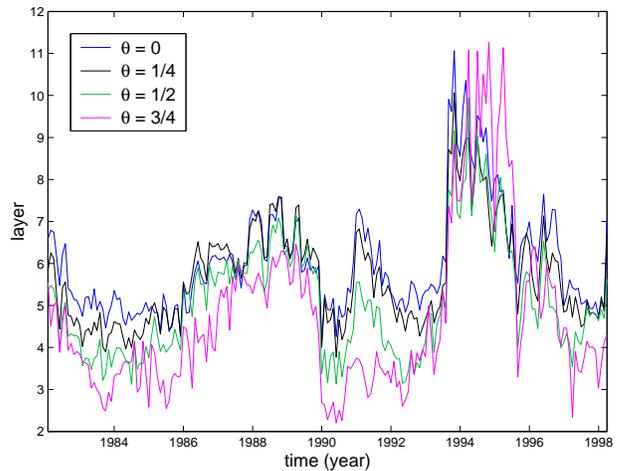,width=3.2in }
\caption{
Plots of the weighted minimum risk portfolio layer
$l_{\mathbf{P}}(t,\theta)$ for different values of $\theta$.
}
\label{port_layers}
\end{figure}

Thus far, we have only examined the location of stocks in the 
minimum risk portfolio, for which $\theta = 0$. However, we note that as we 
increase $\theta$ towards unity, portfolio risk as a function of time soon 
starts behaving very differently from the mean correlation 
coefficient and normalized tree length as shown in Fig. \ref{port_layers}. 
Consequently, it is no longer useful in describing diversification 
potential of the market. However, another interesting result 
is noteworthy: The average weighted portfolio layer 
$l_{\mathbf{P}}(t,\theta)$ decreases for increasing values of 
$\theta$. This implies that out of all the possible Markowitz 
portfolios, the minimum risk portfolio stocks are located 
furthest away from the central vertex, and as we move towards 
portfolios with higher expected return, the stocks included in 
these portfolios are located closer to the central vertex. 
It may be mentioned that we have not included the 
weighted portfolio layer for $\theta=1$, as it is not very 
informative. This is due to the fact that the maximum return 
portfolio comprises only one asset (the maximum return asset 
in the current time window) and, therefore, 
$l_{\mathbf{P}}(t,\theta=1)$ fluctuates wildly as the maximum 
return asset changes over time.

We believe these results to have potential for practical application. 
Stocks included in low risk portfolios are consistently located further away 
from the central node than those included in high risk portfolios. 
Consequently, the radial distance of a node, i.e. its occupation layer, is 
meaningful. We conjecture that the location of a company \emph{within} the 
cluster reflects its position with regard to internal, or cluster 
specific, risk. Thus the characterization of stocks by their branch, 
as well as their location within the branch, would enable us to 
identify the degree of interchangeability of different stocks 
in the portfolio. In most cases, we would be able to pick two stocks from 
different asset tree clusters, but from nearby layers, and interchange 
them in the portfolio without considerably altering the characteristics of the
portfolio. Therefore, dynamic asset trees may facilitate 
\emph{incorporation of subjective judgment} in the portfolio 
optimization problem.

\section{Summary}
We have studied the dynamics of asset 
trees and applied it to market taxonomy and portfolio analysis. 
We have noted that the tree evolves over time and the  mean 
occupation layer fluctuates as a function of time, and experiences 
a downfall at the time of market crisis due to topological 
changes in the asset tree. For the portfolio analysis, 
it was found that the stocks 
included in the minimum risk portfolio tend to lie on the 
outskirts of the asset tree: on average the weighted portfolio 
layer can be almost one and a half levels higher, or further away 
from the central vertex, than the mean occupation layer for window 
width of four years.  Finally, the asset tree can be used 
as a visualization tool, and even though it is strongly 
pruned, it still retains all the essential information of 
the market (starting from the correlations in stock prices) and can be used to 
add subjective judgement to the portfolio optimization problem.

%%%%%%%%%%%%%%%%%%%%%%%%%%%%%%%%%%%%%%%%%%%%%%%%%%%%%%%%%%%%%%%%%%%%%

\section{acknowledgements}
The author would like to thank all his collaborators, and also the critics for 
their valuable comments during the lectures given at IPST (Maryland, USA), 
Bose Institute (Kolkata) and MMV (BHU, Varanasi).

%%%%%%%%%%%%%%%%%%%%%%%%%%%%%%%%%%%%%%%%%%%

\end{document}